\UseRawInputEncoding
\documentclass[twoside]{procinasanen}
\input{procinasan.def}
\usepackage[utf8]{inputenc}
\usepackage{changes}
\usepackage{lineno}
\usepackage{orcidlink}
\newcommand{\msun}{\mbox {$M_{\odot}$}}
\newcommand{\te}{\mbox {$T_{\mathrm{eff}}$}}
 
\newcommand{\ls}{\mbox {$L_{\odot}$}}
\newcommand{\rsun}{\mbox {$R_{\odot}$}}
\newcommand{\mdot}{\mbox {$\dot{M}$}}
\def\apgt{\ {\raise-.5ex\hbox{$\buildrel>\over\sim$}}\ }
\def\aplt{\ {\raise-.5ex\hbox{$\buildrel<\over\sim$}}\ }
\def\fm{\hbox{$.\!\!^{\rm m}$}} 
\def\d{^{d}\kern-2.1mm .\kern+.6mm} 

\begin{document}
\thispagestyle{titlehead}
\setcounter{footnote}{0}
\setcounter{equation}{0}
\setcounter{section}{0}
\setcounter{figure}{0}
\setcounter{table}{0}

\selectlanguage{english}

\markboth{L.R. Yungelson et al. }{Stripped helium stars}

\titlen{Stripped helium stars in UV}
{Yungelson L.R.\orcidlink{0000-0003-2252-430X}$^1$, Kuranov A.G.$^{1,2}$, Mishakina A.V.$^{1,3}$}
{$^1$Institute of Astronomy of the RAS, Moscow, Russia\\
{\normalfont email: lev.yungelson@gmail.com}\\
$^2$Sternberg Astronomical Institute, MSU, Moscow, Russia\\
$^3$Moscow Institute of Physics and Technology, Dolgoprudnyj, Russia}

\abstre{The model of Galactic population of stripped helium stars, the remnants of 
$5 \lesssim  M/M_\odot \lesssim 24$  
primary components of close binaries after RLOF is presented. 
These stars emit in the UV range of the spectrum and may be observed by detectors aboard planned 
``Spektr-UV'' (WSO-UV) space telescope. 
For the current star formation rate in the Galaxy 2$\mathrm {M_\odot yr^{-1}}$, the estimate of the
number of 1 to 7 \msun\ objects with $T_{\rm eff} \gtrsim$25,000\,K ranges from 14700 to 28500. 
The number of 2 to 7 \msun\ objects is 3200 to 5500. 
The main evolutionary factor influencing the estimate is possible formation of common envelopes and merger of components in them. 
Detection of candidate stripped helium stars by UV-excess is hampered by the dominance of main-sequence 
companions to stripped helium stars in the combined spectra of binaries. 
This effect of observational selection may reduce the number of candidate stripped helium stars potentially detectable by photometry to $\lesssim 1000$.
}
\medskip


\medskip


\medskip


\selectlanguage{english}

\baselineskip 12pt
\section*{1. Introduction}
\label{s:intro}
Stripped helium stars (hereinafter referred to as ``helium stars'' or
HeS) are the stars with the masses of approximately (1 -- 7)\,\msun, the remnants of components of 
close binaries (CBS) that lost most of their 
envelopes as a result of Roche lobe overflow shortly before the end of the
main-sequence or during the hydrogen-shell burning stage. 
They are helium cores surrounded by low-mass ($\ll$1\,\msun) hydrogen-helium envelopes.
Possibility of their existence was first demonstrated in the seminal paper by Kippenhahn and Weigert \cite{1967ZA.....65..251K}.

For the metallicity $\rm {Z=Z_\odot}$, the masses of HeS progenitors are 
$M_{1.0}\approx$(5 -- 22)\,\msun.
Stars in this mass range are associated with formation of Be stars, 
progenitors of SN~Ia and SN~Ib/c, Be/X and $\gamma$\,Cas type X-ray sources, single and 
binary neutron stars, sources of gravitational wave signals and still hypothetical 
Thorne-{\.Z}ytkow objects (see, e.g., \cite{2014LRR....17....3P}).
In the Hertzsprung-Russell diagram (HRD), HeS stars are located between hot subdwarfs
(sdB/O) and Wolf-Rayet (WR) stars in the region with
$\log (T_\mathrm{eff})\approx (4.4 - 5.1)$, $\log (L/L_\odot)\approx (2.5 - 5.0)$
(see Fig.~\ref{f:hrd} below).
Due to their high temperatures, they may be important sources of
ionizing radiation \cite{2016MNRAS.456..485S}.
A significant difference between HeS and more massive WR stars should be that their
stellar winds are optically thin and strong emission lines should
be observed only in the spectra of the most massive of them \cite{2018A&A...615A..78G}.

The masses of WR stars exceed 7\,\msun.
There are several thousand of them in the Galaxy
\cite{2015MNRAS.447.2322R, 2020MNRAS.493.1512R}.
Thousands of hot subdwarfs have been discovered. Their masses typically do not exceed
approximately 0.8\,\msun\ \cite{warwick171894,2023ApJ...942..109L}.
However, less than 20 hot subdwarfs with masses $\approx$(1 -- 2)\,\msun\ are known.
There are objects of the similar mass  which are assumed to be in the post-mass-loss stage of evolution
and are contracting, moving to the high-temperature region on the HRD, or have already 
completed core He-burning stage \cite{2020A&A...639L...6S,2022MNRAS.516.3602E,
2023AJ....165..203W, 2025ApJ...995..180G,2018A&A...615A..30S}.

While existence of HeS was predicted about 60 years ago, to the moment only one HeS has 
been detected in the Galaxy -- WR~2-1 \cite{2026arXiv260805276M} with
$\mathrm {T_{eff} = 60^{+3}_{-2}\,kK, \log(L/\ls)= 4.91^{0.05}_{-014}}$.
The mass is estimated as (3.2 -- 5.8)\,\msun.
An interesting binary is $\gamma$\,Col. According to Irrgang et al. 
\cite{2022NatAs...6.1414I},
hot component of the system is observed and its position in the Hertzsprung-Russell diagram
and enhanced abundance of CNO-cycle burning  products at its surface 
suggest that it is an object of several \msun,
transforming into HeS after termination of the  RLOF. 
However, based on the same data, Jin and Langer \cite{2026NatAs.tmp..175J} claim  that the 
hot component of $\gamma$\,Col is, in fact, accreting component of the system.
$\gamma$\,Col is suspected of having pulsations \cite{2006A&A...452..945T}.
The hot component of the binary MWC~656 has mass close to the lower limit of HeS masses 
within the error bars: $M=1.48^{+0.55}_{-0.46}\msun$ \cite{2026A&A...708A.187M}.
Relative to this star, as well as the subdwarf component of the
$\phi$\,Per system, due to their high luminosity and $\mathrm {T_{eff}}$, it is assumed that
they are in the shell He-burning stage 
\cite{2026A&A...708A.187M,2018A&A...615A..30S}.

Stripped helium stars and HeS candidates are found in the Magellanic Clouds
\cite{2023MNRAS.525.5121V,2023ApJ...959..125G,2023Sci...382.1287D,2023A&A...674L..12R,2024A&A...692A..90R,2026ApJ...999...73L}.
Binarity of massive stars is apparently independent of metallicity
\cite{2025NatAs...9.1337S,2025A&A...698A..41V}.
Models of hot post-RLOF remnants of stars in the Magellanic Clouds differ 
little from the models of Galactic stellar remnants: they have a bit more massive envelopes and, as a result, lower \te\ \cite{2024A&A...687A.215D}.
Therefore, it is desirable to consider possible evolutionary reasons for the paucity 
of HeS in the Milky Way and the selection effects that may prevent their detection.

The results of related study are presented in Sections 2 and 3.
Considering the small number of massive subdwarfs observed in the Galaxy, we also classify 
(1 -- 2)\,\msun\ helium remnants of donors in close binaries as HeS.

\section*{2. Population of stripped helium stars}
\label{s:rez}

Computations of the evolution of close binaries resulting in the formation of HeS were performed using  
versions 12778 and r24.03.1 of MESA stellar evolution code (see \cite{2023ApJS..265...15J} and references therein). 

Main assumptions used in the computations are described by Yungelson et al.\cite{2024A&A...683A..37Y,
yungelson2026}.
In  \cite{yungelson2026}, at difference to \cite{2024A&A...683A..37Y}, rotation of both components of 
the system was taken into account.
For  \cite{yungelson2026}, a total of about  1,300 evolutionary tracks were calculated for 
close binaries with initial primary masses 
ranging from 4 to 24\,\msun, orbital periods from 2 to 4000 day, and mass ratios of components from 0.4 to 0.9.
The number of HeS and their distributions over parameters were estimated using population synthesis.
We note only that in the population synthesis in \cite{yungelson2026} a mass-dependent rate of stellar binarity was 
taken  after \cite{2013A&A...552A..69V}, initial mass function of the primary components
followed  Kroupa \cite{2001MNRAS.322..231K}, and current star formation rate in the Galaxy of
2\,\msun\,$\text{yr}^{-1}$ was assumed \cite{2011AJ....142..197C}.
To assess the possibility of merger of  components in common envelopes, energy formalism of Webbink
\cite{1984ApJ...277..355W} and de Kool \cite{1990ApJ...358..189D} was used.
Thai et al.~\cite{2026arXiv260706333T} claim that this formalism describes evolution in the common envelopes best.  
Common envelopes efficiency parameter $\alpha_{\rm CE}$ was set to 1 due to its uncertainty.
Since the issue of release of the internal energy of the donor envelope in the common envelopes remains controversial
\cite{2026arXiv260706333T,2020cee..book.....I},
the primary scenario was the case in which it was assumed that the internal
energy is completely spent on envelope removal
\footnote{Test calculations have shown that, when
considering only contribution of the binary orbital gravitational energy,
the results do not change, since components in both cases merge in the common envelopes.
}.
Donor envelope binding energy parameter, necessary for estimating the change in the distance
between the components in the case of common envelope formation, was calculated for every model.
Models of the remnants of stars with $1 \leq M/M_\odot \leq 7$ and $\te \geq 25000$\,K were considered as HeS.

\begin{figure}[t] 
\begin{center}
\includegraphics[width=0.6\textwidth]{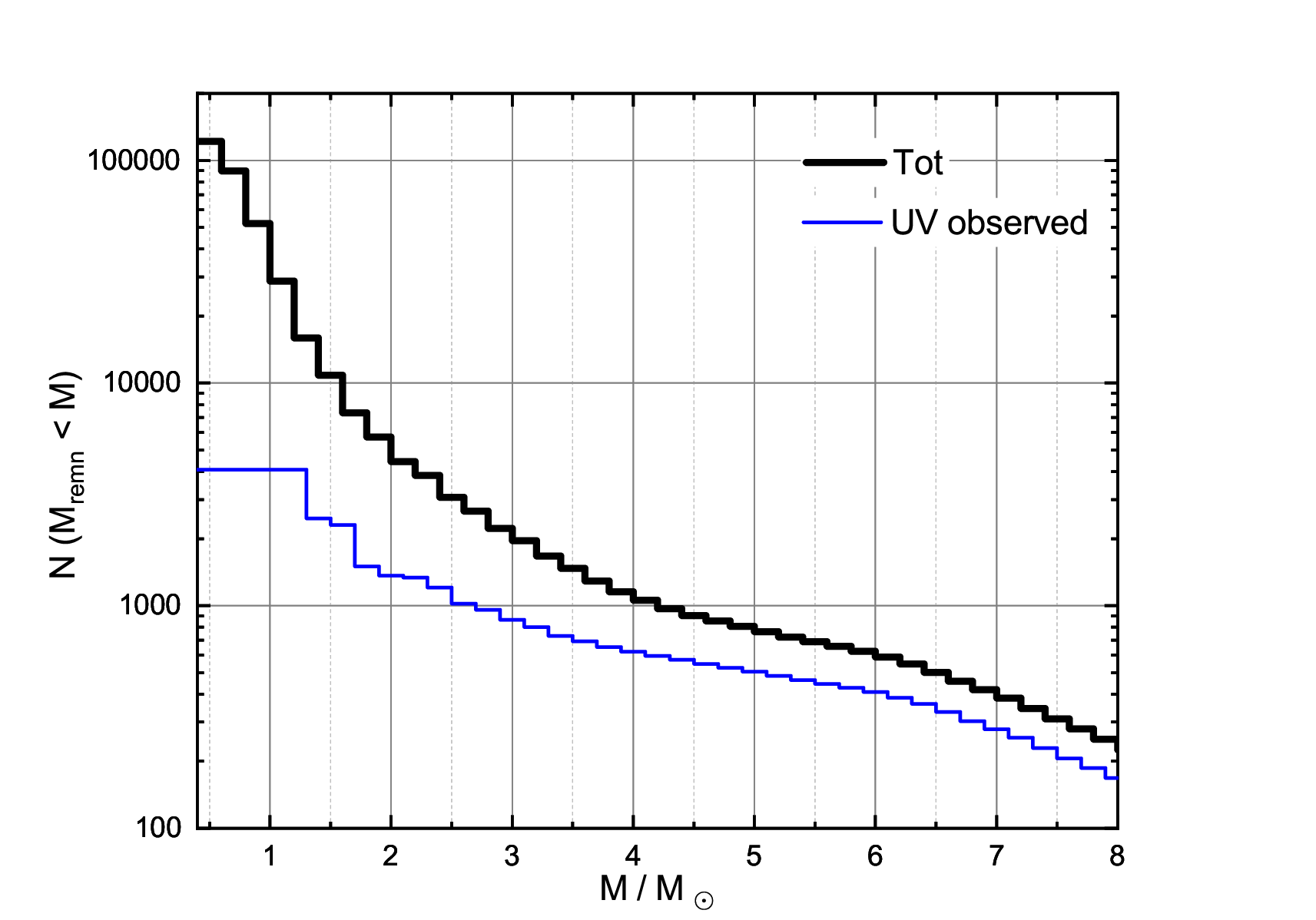}
\caption{Cumulative distribution of the number of remnants of donors in close binaries at 
the core He-burning stage.
As HeS are considered the remnants  with $M_{\rm remn} \geq$ 1\,\msun.
Black line shows the total number of stars, blue line shows the number of stars,
detection of which is possible by UV excess (Section~3).
}
\label{f:number}
\end{center}
\end{figure}

Figure \ref{f:number} shows the cumulative distribution  of the number of helium remnants of the primaries 
of close binaries during core He-burning stage.
Formation of the population of HeS with main-sequence companions involves close binaries with the 
primaries overflowing Roche lobes either on the main-sequence and in the H-shell burning stage 
(cases A and B  of mass exchange, see \cite{2014LRR....17....3P}).
Total Galactic population of HeS is estimated as 28,500 objects. Only 5,500 stars have masses exceeding 
2\,\msun.
About 10\% of HeS are formed by systems with RLOF in case A, the rest comes from case B systems.
Relative paucity of the former is due to the fact that, because of increasing radius of the accretor,
contact systems are formed, and the components merge when the envelope
reaches the second Lagrange point, from which the matter is lost with a significant angular momentum.
For close binary systems with RLOF in the Hertzsprung gap, the efficiency of HeS formation
is limited primarily by the unstable mass loss and formation of the common envelopes.

\begin{figure}[h!] 
\begin{center}
\includegraphics[width=0.6\textwidth]{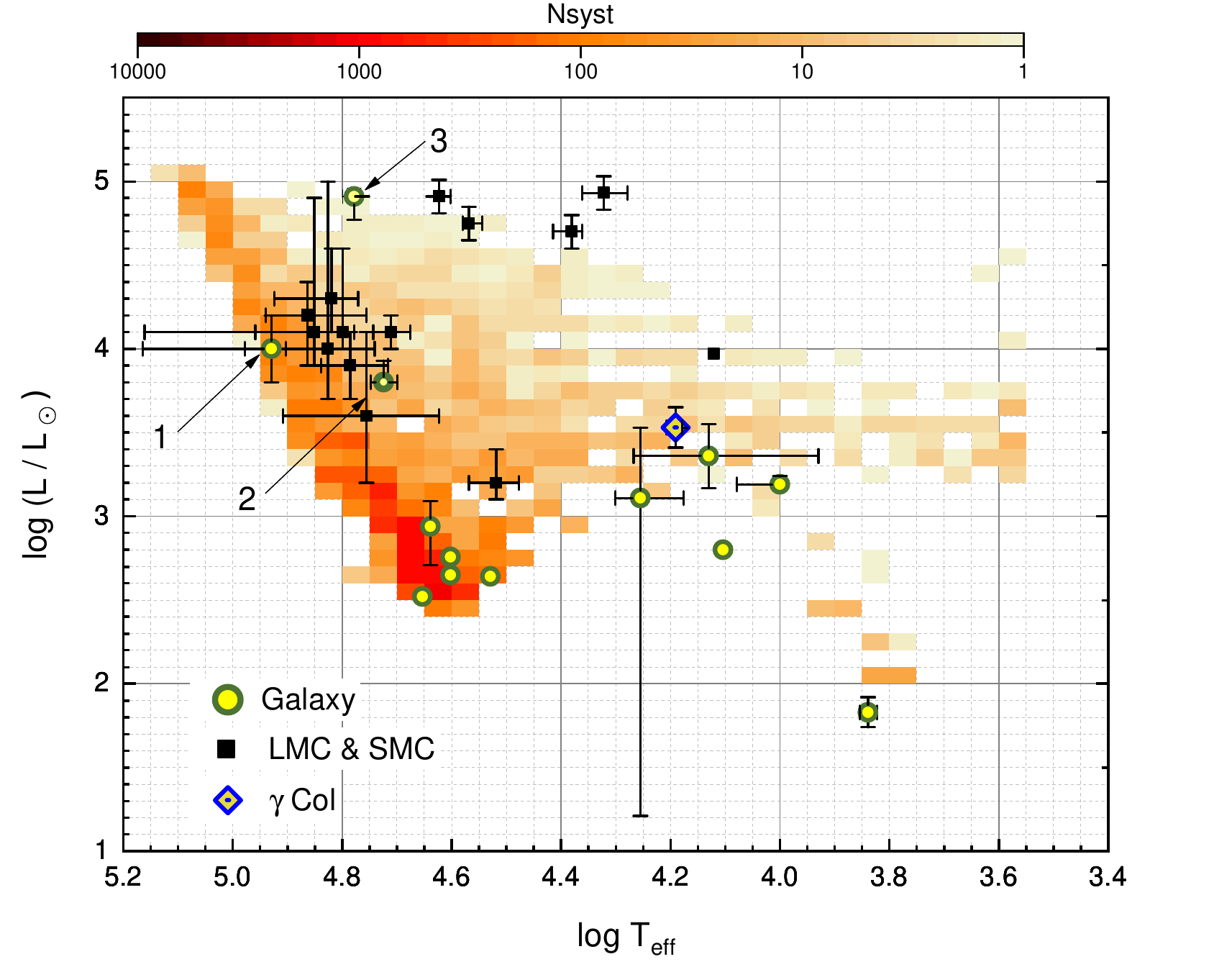}
\caption{Distribution of the number of HeS and their precursors in the Hertzsprung-Russell diagram.
Yellow symbols are massive ($M \apgt$1\msun) helium subdwarfs of the Galaxy.
Black symbols are stars found in the Magellanic Clouds.
Position of $\gamma$\,Col is marked by a diamond.
1 -- MWC~656,
2 -- $\phi$\,Per,
3 -- WR2-1.
}
\label{f:hrd}
\end{center}
\end{figure}
\begin{figure}[b!] 
\begin{center}
\includegraphics[width=0.6\textwidth]{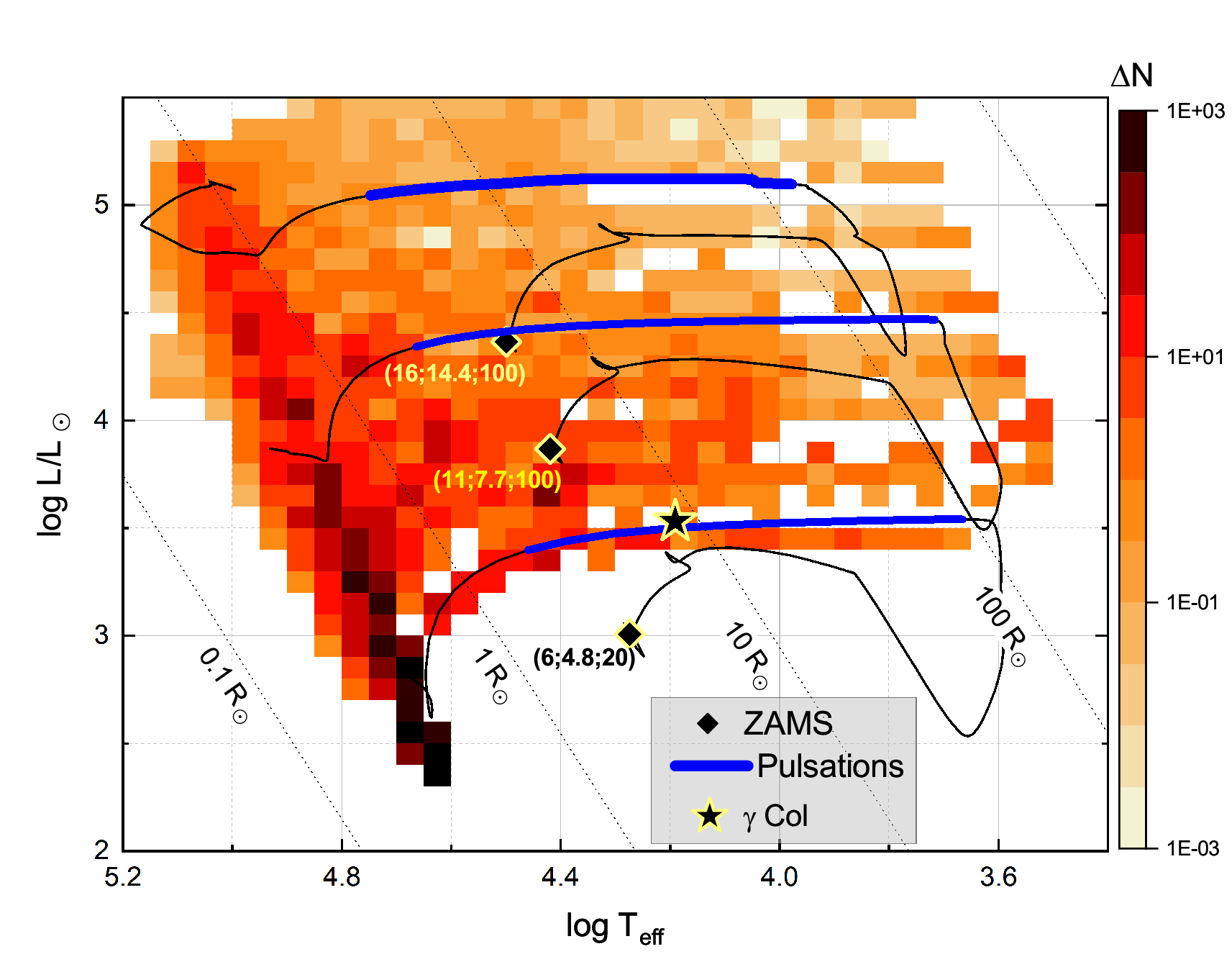}
\caption{Distribution of the number of HeS and their precursors in the Hertzsprung-Russell diagram
(color-coded)
\cite{2024A&A...683A..37Y}.
Tracks of the primary components of the close binaries with masses 9, 12, 16\,\msun\ are shown.
The sections of the tracks on which HeS precursors pulsate according to \cite{2025AstL...51...25F} are highlighted in blue.
Position of $\gamma$\,Col is marked.
}
\label{f:fad_hrd}
\end{center}
\end{figure}

Figure~\ref{f:hrd} shows model distribution of the number of HeS and their contracting precursors in HRD.
Positions of WR2-1, $\gamma$\,Col, $\phi$\,Per and known massive Galactic helium subdwarfs as well 
as  HeS and the candidates found in the Magellanic Clouds are shown.
As mentioned in the Introduction,
HeS in the MC may be somewhat ``cooler'' than predicted for HeS in the Milky Way.
Simulation results, at least qualitatively, reproduce observations well.

Hovis-Afflerbach et al. \cite{2025A&A...697A.239H}, in addition to RLOF, considered possibility of 
formation of HeS via common envelopes and
estimated the fraction of HeS formed via this channel as $\simeq 20$\%.
However, this fraction depends on the adopted parameters of the common envelope efficiency
$\alpha_{\rm CE}$ and binding energy of the donor envelope $\lambda$.
In \cite{2025A&A...697A.239H}, these parameters were set to 1 and 0.5, respectively.
But if in the present study $\alpha_{\rm CE}$ was also assumed to be equal to 1,  $\lambda$ was calculated 
using stellar models and, as a rule, $\lambda \ll 0.5$.
Since, roughly, the distance between components of a close binary after the end of the common envelope 
stage is proportional to $\alpha_{\rm CE} \times \lambda$,
under assumptions adopted in this paper, evolution in the common envelopes always results in  the merger
 of components.

Zapartas et al. \cite{2017ApJ...842..125Z} called attention to the possibility of formation
of HeS in the binaries with  compact objects where Roche lobe is overflown by the  the secondary component.
Estimate of the number of such systems depends on the assumptions regarding
efficiency of accretion, common envelopes and characteristics of supernova  explosion that formed compact object.
Such helium stars should not be ``outshined'' by companions and may be disproportionately well
represented in the observed samples \cite{2026PASP..138b4202B}.

Picco et al. \cite{picco2026mergers} suggested a scenario in which a {\it single} HeS
forms as a result of the merger of binary components in a common envelope.
It is assumed that the system already hosts a HeS and the Roche lobe is filled by its evolved companion.
This scenario may occur only in the tightest binaries with initial $M_2/M_1 \apgt 0.75$ and orbital
periods $P \aplt 30$\,day. The likelihood of this scenario depends on the assumed 
parameters of common envelopes and convective mixing.
As Picco et al. suggest, this scenario can explain formation of the hot component
of wide binary system HD~45166, a quasi-WR (qWR) star \cite{2025A&A...695L..20D}.
It is assumed that the qWR star is a product of merger in a close binary
in an initially triple system of a close system and a distant companion.
The mass of the qWR star is $1.96_{-0.54}^{+0.74}$\,\msun.
If its mass really exceeds about 2.5\,\msun, it should remain
compact after helium exhaustion in the core \cite{1970AcA....20..195P},
its mass will not decrease significantly due to the stellar wind and
it can explode as a supernova producing a magnetar \cite{2025A&A...697A.101L}.

Immediate progenitors of HeS may represent  a new type of variable stars.
Fadeyev et al. \cite{2025AstL...51...25F} have shown that at the stage
of contraction after cessation of RLOF the stars pulsate non-radially with 
bolometric amplitudes up to 0\fm 8 and periods from 0.17 to 8.9 day.
Figure~\ref{f:fad_hrd} shows the distribution of HeS number over HRD
\cite{2024A&A...683A..37Y} and evolutionary tracks of the primary components of the close binaries
 with initial donor masses 9, 12 and 16\,\msun.
Sections of the tracks where stars pulsate are marked.
Pulsation amplitudes increase with decreasing stellar radii.
Position of the star $\gamma$\,Col is shown, which, as already
noted, is suspected to
pulsate \cite{2006A&A...452..945T}\footnote{Assuming that it is really former donor in the system 
\cite{2022NatAs...6.1414I}.}.
The estimate of the number of pulsating pre-HeS in the Galaxy in a model in which
accretion is rotation-limited is $\sim$1000.
However, formation of the common envelopes  due to impossibility to remove 
non-accreted matter from the system can significantly reduce their number, while 
observational selection makes it difficult to detect them (see below).

{\it An evolutionary effect} possibly limiting the number of HeS, may be formation of the 
common envelopes  and merger of binary components in them.
Common envelopes may result from the unstable mass loss by the 
donors at a high rate, which is taken into account in the estimates presented above.
In the model with rotation-limited accreton rotation of the secondary component becomes critical after accretion
of only few percent of the matter lost by the donor \cite{1981A&A...102...17P}. 
It is assumed that further accretion is limited by maintaining  critical angular velocity.
Excess of the matter may accumulate in the system. 
As a mechanism that may remove non-accreted matter, as a rule,
radiation pressure of both components is considered.
Excess of the matter leaves the system taking away specific angular momentum of accretor.
If one assumes that the matter leaves the system at the distance from the accretor with mass $M_{\rm 2}$ 
equal to the radius of its Roche lobe $R_{\rm L,2}$, then the maximum rate of mass removal is 
\begin{equation}
\label{eq:mlossmax}
\dot{M}_{exc} \approx 10^{-7.2} \frac{(L_1+L_2)}{\ls} \frac{R_{\rm L,2}}{\rsun} 
\frac{M_\odot}{M_2} \quad \frac{M_\odot}{\text{yr}}, 
\end{equation}       
where $L_1$ and $L_2$ are luminosities of components \cite{2017PhDT.......434M}.
This hypothetical mechanism has not yet been studied.
Equation (\ref{eq:mlossmax}) can contain either current parameters of components or their mass and luminosity at
the end of the mass exchange stage. Latter case is more favourable for removal of the matter.
It is possible that  \mdot\ exceeds $\dot{M}_{exc}$ while the star loses only a fraction of its envelope.
In general, it is unclear what is the distance relative to the accretor at which the matter is removed, 
what fraction of the radiation of the components can be spent in this case, what energy has the 
matter which can be considered as deleted and what is the system configuration if $\mdot > \dot{M}_{exc}$ 
\cite{2024A&A...682A.169H}.
It is possible that a contact system forms,
circumstellar disk or a disk around the system can form, formation of a common envelope is possible.
If a contact system arises and matter is lost from the vicinity
of the second Lagrangian point ${\rm L_2}$, binary probably also plunges into the common envelope,
after a stage of the (over)contact system.
As an extreme case, we can assume that the system plunges into a common envelope
and to compute the evolution in it using energy formalism and
substituting into (\ref{eq:mlossmax}) the current parameters of the stars.
Like in the case of common envelope associated with unstable mass loss, components
of almost all close binaries merge. As a result, the number of HeS formed in case A of mass exchange
decreases to 1700, while in the case B -- to 13000. Thus, the effect
unremovable non-accreted matter  can reduce the total number of HeS by almost 50\%. 
Mostly, close binaries  with initial $M_{2.0}/M_{1.0} \aplt (0.6 - 0.7)$ and
$M_{1,0} \aplt (14 - 18)\,\msun$ become lost,  i.e. precursors of HeS of small and moderate masses.

\section*{3. Observations of stripped helium stars}
\label{s:selection}
\begin{figure}[t!] 
\begin{center}
\includegraphics[width=0.4\textwidth,angle=-90]{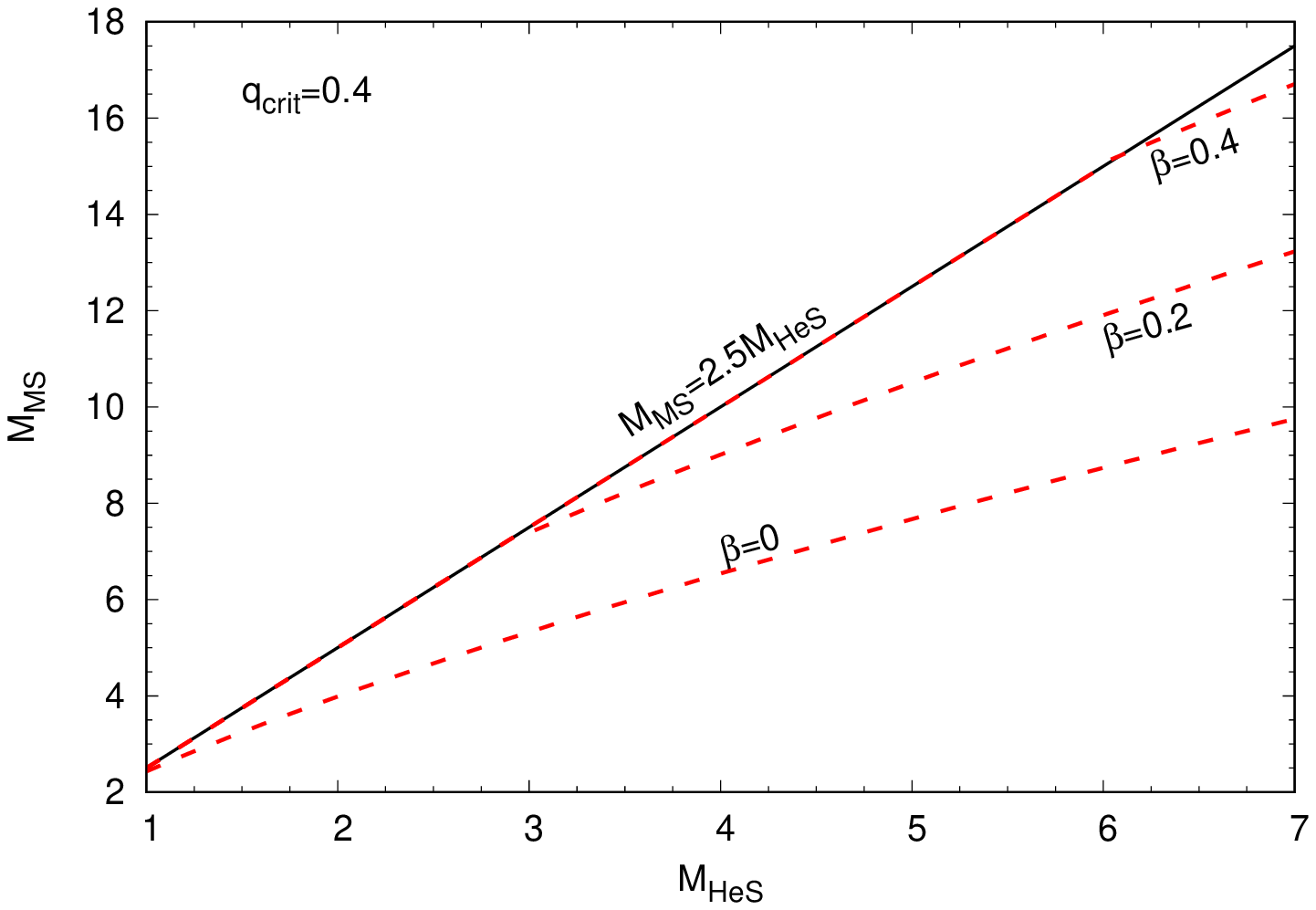}
\caption{Relationship between the masses of components in close binaris harbouring stripped helium stars with main-sequence companions.
If component masses correspond to the part of the diagram above the line $M_{MS} = 2.5M_{HeS}$,
detection of HeS by the UV excess is hardly possible.
Red dashed lines delimit from below the areas where 
HeS can be observed for different values of the accretion efficiency parameter $\beta$.
}
\label{f:m10m20}
\end{center}
\end{figure}

The models of G{\"o}tberg et al. \cite{2018A&A...615A..78G} suggest that  detectable emission 
of HeS is in the wavelength range from a thousand to several thousand \AA, within the sensitivity bands
of the receivers on board of the planned ``Spektr-UV'' (WSO-UV) spacecraft (1500 -- 8000) \AA. 
Since the real possibility of observing Galactic
HeS has not yet been investigated, HeS of the Magellanic Clouds can be used for estimates.

Most numerous HeS candidates in the Magellanic Clouds were identified by
Drout et al. \cite{2023Sci...382.1287D} and Ludwig et al.
\cite{2026ApJ...999...73L} photometrically, based on their positions in the colour-luminosity diagrams.
For a very rough estimate of the upper limit on the number of HeS potentially
detectable in the Galaxy by photometry we can use the criteria found by G{\"o}tberg et al. 
\cite{2018A&A...615A..78G} and
Drout et al. \cite{2023Sci...382.1287D}, according to which HeS can be detected in UV
 if the mass of its main-sequence companon $M_{\rm MS} \aplt (1.7 \div 2.5) M_{\rm HeS}$.
Taking the limit 2.5 and assuming that the same condition is satisfied for Galactic stars, one may 
obtain the following estimate.
For intermediate mass close binaries we can adopt relation between the masses of HeS
and the masses of progenitors:
$M_{\rm HeS}/\msun \approx 0.08(M_{1.0}/\msun)^{1.4}$ \cite{1985ApJS...58..661I}.
Let introduce accretion efficiency parameter:
$\beta = (M_{\rm MS} - M_{\rm 2.0}) / (M_{\rm 1.0} - M_{\rm HeS}).$
Then the  condition for detection is
\begin{equation}
\beta \left [\left ( M_{\rm HeS}/0.08 \right )^\alpha - M_{\rm HeS} \right ] 
+ q_0 \left ( M_{\rm HeS}/0.08 \right )^\alpha \aplt 2.5 M_{\rm HeS},
\label{eq:hes}
\end{equation}
where $\alpha = 1/1.4$, $q_0 = M_{2,0}/M_{1,0}$ -- initial mass ratio of components.
To avoid formation of a common envelope as a result of unstable mass loss by the donor, 
$q_0$ has to be greater than a certain limit $q_{\rm crit}$.
According to our computations, $q_{\rm crit}$ is close to 0.4.
Substituting $q_{\rm crit}$=0.4 into Eq.~(\ref{eq:hes}), we can
identify combinations of masses of components in the HeS+MS systems 
which allow detection of HeS for  different values of $ \beta $ (Fig.~\ref{f:m10m20}).
``Observable'' stars in the $\mathrm {M_{HeS} - M_{MS}}$ diagram are located below the line
$\mathrm {M_{MS}=2.5M_{HeS}}$ and above the lines corresponding to the various $\beta$ in the inequality (\ref{eq:hes}).

As accretion efficiency increases, HeS companions become more massive,
limiting the number of HeS potentially detectable by their UV emission.
In the most favourable case of almost completely non-conservative accretion,
the number of HeS+MS pairs in the ``observable'' region is $\approx 3000$
(see Fig.~\ref{f:number} above).
There are only $\approx 1000$ stars with masses $\apgt 2$\,\msun.

Clearly, only non-abundant stars with $M_{\rm HeS} \simeq$(2 -- 4)\,\msun\ (Fig.~\ref{f:number}) should be detectable.
The number of the most massive HeS stars should be small due to the initial mass function.
In the case of non-conservative matter exchange, leading to an increase in the accretor mass, the ``observable region'' will become even smaller.
For $\beta \apgt$0.4, HeS stars potentially detectable by their UV emission virtually disappear.

If we would take  for the estimate the  limit of  mass ratio of components
$M_{\rm MS} \aplt 1.7M_{\rm HeS}$, the number of ``detectable'' HeS stars will also decrease.
On the other hand, $q_{\rm crit}$ may depend on the stellar masses and be lower than 0.4,
while the accretion may be conservative to certain extent.
However, ``outshining'' of HeS by their companions, the main-sequence stars, in UV  may be the main
{\it observational selection effect} that prevents HeS from being detected by photometry.

To estimate the number of HeS actually accessible for observations in the Galaxy, it is necessary to 
construct combined spectra of HeS  and their companions, taking into account their age and possible 
accretion effects, to distribute them randomly over Galactic disk and to take into account  extinction and 
the sensitivity of the detectors.
Furthermore, as Ludwig et al. noted in their analysis of observations
of HeS candidates in the Magellanic Clouds \cite{2026ApJ...999...73L}, UV color excesses similar to those of HeS stars may be found, 
in particular, in early-type main-sequence stars, RR Lyrae stars, and still hypothetical so-called 
``Rapidly accreting white dwarfs'' (RAWD, \cite{2013ApJ...771...13L}), hot subdwarfs, and stars that have 
completed their evolution on the asymptotic
giant branch.``Filters'' or excess thresholds must be introduced to separate the latter stars and to 
eliminate possible blending. Thus, the number of ``observable'' objects may be $\ll 1000$.
Blomberg et al. \cite{2026ApJ...999...73L} simulated the population of HeS in the Magellanic Clouds
and the search for them in the Stripped-Star Ultraviolet Magellanic Cloud Survey,
carried out by Ludwig et al. \cite{2026ApJ...999...73L}, accounting for selection effects.
Estimated helium star  detection efficiency, if using UV photometric data obtained by the Swift 
spacecraft's UVOT instrument \cite{2005SSRv..120...95R}, does not exceed 10\%.
Note, both WR2-1 and the stars with known parameters found in the Magellanic Clouds 
\cite{2023MNRAS.525.5121V, 2023A&A...674L..12R, 2024A&A...692A..90R} have been identified spectroscopically.
The candidates found in \cite{2023Sci...382.1287D, 2026ApJ...999...73L} are awaiting confirmation.

\section*{4. Conclusion}

The analysis of the formation of the Galactic population of stripped helium stars and the possibility of 
detecting them allows us to draw the following conclusions.

Stripped helium stars certainly exist, as predicted by the stellar evolution theory almost  60 years ago. 
There is at least one candidate star in the Galaxy and they have been discovered in the Magellanic Clouds.

Stripped helium stars have $\te  \apgt$50,000\,K  and the maximum of their emission fluxes
is in the ultraviolet part of the spectrum detectable by ``Spektr-UV'' (WSO-UV).

Estimated population of HeS with masses from 1 to 7 \msun\ may reach 28,500.
However, if the accretion rate is limited after the accretor attains critical rotation velocity and 
non-accreted mass accumulates in the binary system, possibility of the formation of common envelopes 
where components of the systems merge, cannot be ruled out. In this case, HeS population in the 
close binary stars could be reduced by a half, to 14,700 objects.
Mergers could result in the formation of isolated HeS. 
The most important effect of the observational selection preventing the detection of
HeS stars could be the ``outshining'' of HeS stars by their main-sequence companions, if
$M_{\rm MS} \apgt 2.5M_{\rm HeS}$.
This effect may further severely reduce the number of potentially observable HeS. 

Further selection could be related to the distance to HeS and interstellar absorption.
Selection will be more effective if the loss of matter by the donor in the close binary star is unstable
for the initial mass ratio of components exceeding $q_{\rm crit}= 0.4$, as assumed by us or if the mass 
exchange is more conservative than in the model with rotation-limited accretion. 

\vspace{0.1cm}
The authors acknowledge A.M. Cherepashchuk, K.A. Postnov, L. Piersanti,
I.A. Shaposhnikov, and L.I. Mashonkina for helpful discussions.
This work was supported by Russian Science Foundation grant No. 25-22-00295, https://rscf.ru/project/25-22-00295/.
NASA ADS bibliographic database has been used in this work.

\bibliographystyle{inasan}
\small
\bibliography{./yungelson}
\normalsize
\end{document}